\documentclass[9pt,conference]{IEEEtran}
\usepackage{dcase2026}

\usepackage{bm} 

\usepackage{dcase2026,amsmath,graphicx,url,times,booktabs, tabularx}

\usepackage{multirow}
\usepackage[table]{xcolor}

\title{SAS: Scene-Annotated AudioSet Strong for Joint Scene-Event Analysis}

\name{Michel Olvera$^{1}$,
      Paraskevas Stamatiadis$^{1}$,
      Romain Serizel$^{2}$,
      Slim Essid$^{3*}$\thanks{*Work conducted while at Télécom Paris.},
      Gaël Richard$^{1}$}
\address{\normalsize$^{1}$LTCI, Télécom Paris, Institut Polytechnique de Paris, France \;
$^{2}$Université de Lorraine, CNRS, Inria, Loria, France \;
$^{3}$NVIDIA
}

\begin{document}

\maketitle

\begin{abstract}
Context is central to human auditory perception, as listeners interpret sounds within their acoustic environment to resolve ambiguity. However, large-scale datasets that jointly capture sound events and acoustic scenes remain limited. We address this by augmenting AudioSet Strong with clip-level scene annotations under two taxonomies for joint scene-event modeling. A corpus-scale analysis shows that scene identity is better captured by events distinctive to a scene than by those that merely occur frequently. Reweighting event co-occurrences toward scene-specific associations produces representations that better align with human-derived taxonomies, while also exposing scenes where the acoustic and semantic views diverge. This geometric divergence predicts classification errors, while supervised validation confirms that the proposed scene annotations provide meaningful complementary information for developing scalable context-aware machine listening systems.
\end{abstract}

\begin{IEEEkeywords}
Acoustic scene classification, sound event detection, scene-event modeling, context-aware sound recognition, sound semantics
\end{IEEEkeywords}

\section{Introduction}
\label{sec:intro}

Everyday sounds are interpreted in context: a \emph{siren} wails on a \emph{street} but jars in a \emph{library}, and a \emph{kettle}'s whistle is unremarkable in a \emph{kitchen} but anomalous in a \emph{garage}. Human listeners use the recognized environment to anticipate likely sound sources, suppress implausible ones, and resolve acoustic ambiguities \cite{ballas1991effects, heittola2013context}. Reproducing this capability is the goal of \emph{context-aware machine listening}, in which systems go beyond sound recognition to jointly infer the \emph{sound event} and the surrounding \emph{acoustic scene}, so that each constrains and disambiguates the other. Although this premise is well established, datasets and models that jointly leverage it at scale remain scarce.

Research on scene-event relationships has followed two main directions. Multi-task learning approaches jointly optimize scene and event recognition by exploiting shared information, typically through implicit representations of their dependencies \cite{tonami2019joint, imoto2020sound, igarashi2023joint, krause2022binaural}. These efforts have motivated the creation of datasets with scene and event labels \cite{mesaros2016tut, tonami2019joint, martin2023strong}, but such resources remain limited in scale, covering only a small number of scenes, events, and recordings. Conversely, large-scale weakly labeled resources such as AudioSet \cite{gemmeke2017audio} have enabled event modeling across millions of recordings and hundreds of classes, while AudioSet Strong \cite{hershey2021benefit} adds temporally localized annotations for a subset of this data.

However, neither AudioSet nor AudioSet Strong includes scene annotations, so the event and scene ontologies they support remain disconnected, and the statistical structure linking acoustic scenes and sound events remains uncharacterized at corpus scale. Some sound events may provide strong contextual evidence because they are highly specific to particular environments, while others may primarily reflect generic acoustic properties shared across many scenes. Without large-scale data with both event and scene labels, it is not possible to test which of these regimes predominates, or, more fundamentally, whether the acoustic identity of a scene is better predicted by the events that occur most frequently within it or by the events that are most informative in distinguishing it from other scenes.

\begin{figure}[t]
  \centering
  \centerline{\includegraphics[width=\columnwidth]{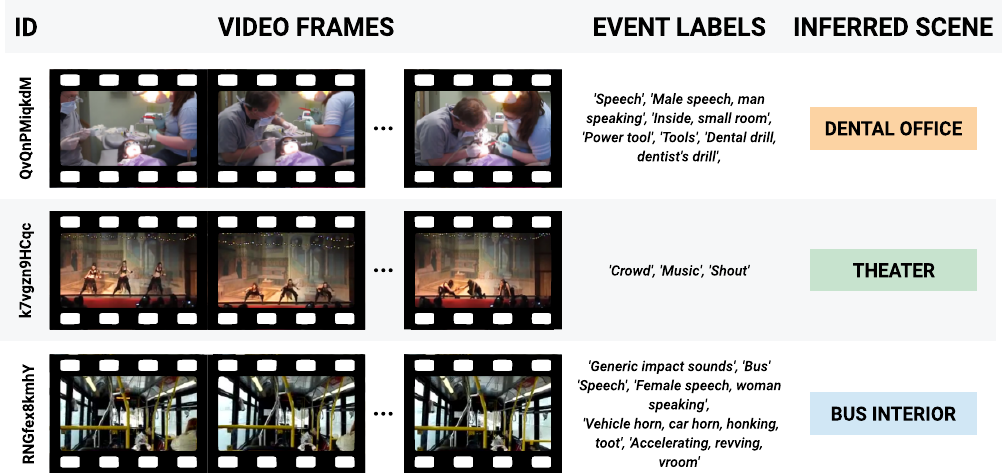}}
\caption{We propose extending AudioSet Strong with scene labels to foster joint analysis of sound event and acoustic environment relationships.}

  \label{fig:teaser}
\end{figure}

To make this contextual information explicit, we augment AudioSet Strong with acoustic scene labels\footnote{\url{https://github.com/michelolzam/scenes-audioset}} (\cref{fig:teaser}). The resulting resource provides clip-level scene annotations for 207,315 unique video identifiers, supporting analysis of scene–event relationships at a scale and coverage not available in prior scene-event corpora. Scene labels are organized under two proposed complementary scene taxonomies, one audiovisual and one audio-centric, allowing scene–event associations to be analyzed at different levels of semantic abstraction. We leverage this resource to conduct a model-agnostic, interpretable analysis of scene-event co-occurrence, testing whether scene identity is driven by event frequency or by event discriminativeness.

Overall, this work introduces three key contributions: (1) an extension of AudioSet Strong with clip-level acoustic scene annotations under two complementary taxonomies; (2) a corpus-scale analysis revealing how sound events distinguish acoustic scenes and explain inter-scene similarity; and (3) a diagnostic validation suggesting that the proposed annotations encode meaningful scene information for joint scene-event modeling.

\begin{figure*}[t]
  \centering
  \includegraphics[width=0.9\textwidth]{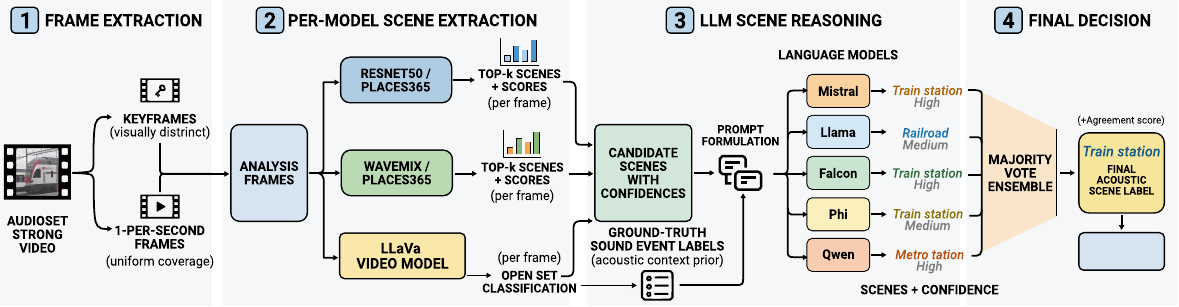}
  \caption{Overview of the acoustic scene label extraction pipeline: visual evidence extraction, cross-modal reasoning, ensemble voting, and taxonomy mapping.}
  \label{fig:pipeline}
\end{figure*}

\section{Acoustic Scene Labeling Pipeline}
We propose a four-stage cross-modal pipeline (\cref{fig:pipeline}) that assigns global acoustic scene labels to audio clips by integrating visual proxies, sound-event priors, and multi-model late fusion.

\noindent\textbf{Stage 1: Multi-rate visual frame extraction.} Each video is decoded into two complementary visual proxy sets: (i) uniform sampling (1 Hz), following the temporal resolution used in strongly annotated sound event datasets \cite{martin2023strong}, and (ii) I-frame samples to capture content transitions. The frame budget is limited to ten uniformly sampled frames and up to ten keyframes, with the same frames provided to all visual models.

\noindent\textbf{Stage 2: Scene candidate generation.}
Three vision models with complementary label spaces and prediction paradigms process the extracted frames:

\begin{itemize}
\item Two closed-set image classifiers (ResNet50 \cite{he2016deep} and WaveMix \cite{jeevan2022wavemix}), pretrained on the Places365 taxonomy, produce frame-level top-5 scene predictions with confidence scores.
\item An open-vocabulary video-language model (LLaVA \cite{liu2023llava}) processes the frames under two prompt variants: a \textit{standard} prompt for scene identification and an \textit{uncertainty-aware} prompt that flags uncertain predictions.
\end{itemize}

\noindent\textbf{Stage 3: Cross-modal evidence gathering and LLM reasoning.}
Frame-level outputs are aggregated into clip-level scene candidates. ResNet50 and WaveMix provide closed-set scene predictions with confidence scores, which are averaged across frames to reduce the influence of outlier frames that may not reflect the actual scene (e.g., transient static images at the beginning or end of clips). In contrast, LLaVA produces one open-vocabulary scene prediction per frame, with confidence estimated as the frequency of each prediction across frames. AudioSet Strong event annotations are also aggregated into clip-level event sets and used as acoustic evidence for scene inference.

For each clip, aggregated evidence---four vision predictions (LLaVA with a standard prompt, LLaVA with an uncertainty-aware prompt, ResNet50, and WaveMix), with associated scene candidates and confidence scores, together with the clip-level sound-event set---is formatted into a shared prompt template. Five instruction-tuned language models are used (Mistral 7B \cite{Jiang2023Mistral7}, Llama \cite{grattafiori2024llama}, Falcon \cite{almazrouei2023falcon}, Phi \cite{abdin2024phi}, and Qwen3 \cite{yang2025qwen3}). Model outputs are normalized and mapped to a manually curated audiovisual (AV) taxonomy, then projected onto an audio-centric (AC) taxonomy that aggregates scene labels from the audiovisual taxonomy. The audio-centric taxonomy accounts for cases where visually distinct scenes are acoustically ambiguous, aggregating categories such as \textit{marsh}, \textit{swamp}, and \textit{creek} into broader acoustic environments, e.g., \textit{wetland}.

\noindent\textbf{Stage 4: Final decision by voting.} The final acoustic scene label is obtained through majority voting across the ensemble. Ties are retained as multiple candidates. The released annotations include the vote count and aggregated confidence, obtained by mapping categorical confidence levels (e.g., \textit{low/medium/high}) to a numeric scale. The resulting scene labels provide the basis for the subsequent statistical analysis of scene-event relationships.

\section{Statistical Analysis of Scenes and Events}
\label{sec:analysis}

Our framework assumes that acoustic events encode two complementary aspects of scene identity: their occurrence frequency within a scene and their specificity relative to the corpus. Let $\mathcal{E} = \{e_1, \dots, e_M\}$ be the sound event classes, with $e \in \mathcal{E}$, and $S = \{s_1, \dots, s_K\}$ be the scene labels, with $s \in S$. Let $C \in \mathbb{R}^{M \times K}_{\geq 0}$ be the corpus-wide scene-event co-occurrence count matrix, computed from the training split.

\subsection{Structural views of scene identity}

We derive two complementary scene representations from $C$.

\textbf{1. Acoustic profile view ($\mathbf{p}_s$).} We define it as the empirical conditional distribution of sound events given a scene:
\begin{equation}
    p(e \mid s) = \frac{c(e,s)}{\sum_{e'} c(e', s)}.
\end{equation}

\begin{figure*}[t]
  \centering
  \includegraphics[width=0.98\textwidth]{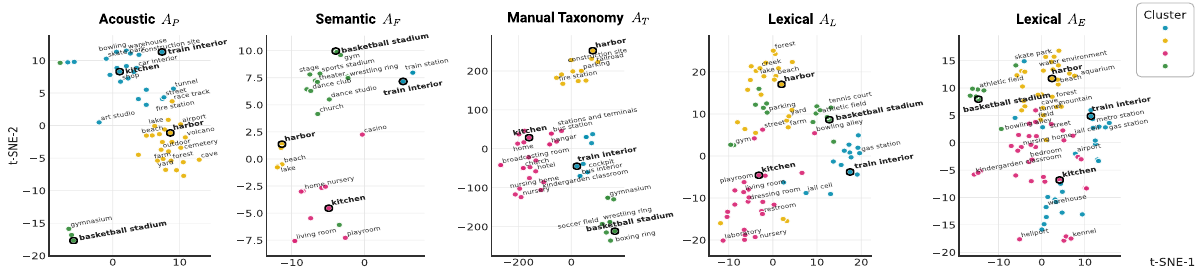}
\caption{2-D projections of representational views with highlighted neighborhoods of four scenes across acoustic, semantic, and lexical spaces.}

  \label{fig:embeddings}
\end{figure*}

The resulting probability vector $\mathbf{p}_s \in \Delta^{M-1}$ captures the sound events present in scene $s$. Since it is dominated by globally frequent events, this representation emphasizes shared background acoustics over scene-specific semantic content.

\textbf{2. Semantic profile view ($\mathbf{f}_s$).} To emphasize scene-specific events over ubiquitous background sounds, we apply Pointwise Mutual Information (PMI) to quantify scene-event association. Let $P(e,s) = c(e,s) / \sum_{e',s'} c(e',s')$ denote the joint empirical probability, with marginals $P(e)$ and $P(s)$. The PMI score is $\mathrm{PMI}(e,s)=\log \frac{p(e\mid s)}{P(e)}$. We retain only positive associations using PPMI and introduce a frequency discount $w(e,s) = \frac{c(e,s)}{c(e,s) + 1}$ to reduce the influence of rare events. The discounted PPMI ($\mathrm{dPPMI}$) score is:
\begin{equation}
\mathrm{dPPMI}(e,s) = w(e,s) \cdot \max\left(0, \,\mathrm{PMI}(e,s)\right).
\end{equation}
Normalizing the resulting profile across the vocabulary gives the semantic event vector $\mathbf{f}_s$:
\begin{equation}
f_s(e) = \frac{\mathrm{dPPMI}(e,s)}{\sum_{e'} \mathrm{dPPMI}(e', s)}.
\end{equation}

\textbf{3. Text embedding views ($\mathbf{b}_s$).}
To compare the previous representations with language-based scene semantics, we use a Sentence-BERT ($\mathtt{SBERT}$) encoder as a text baseline. For each scene label we compute two embeddings:
\begin{equation}
\mathbf{b}_s^{\mathrm{raw}} = \mathtt{SBERT}(s), \qquad \mathbf{b}_s^{\mathrm{prompt}} = \mathtt{SBERT}\!\left(T(s)\right)
\end{equation}
where $T(\cdot)$ is a fixed prompt template that converts a scene label into a natural-language description (e.g., \textit{``The acoustic environment and ambient sounds of a \{scene\}.''}). The raw embedding captures lexical scene semantics, while the semantic prompted embedding incorporates environmental context through the same encoder.

\textbf{4. Manual taxonomy view.} Both our AV and AC taxonomies support multiple inheritance and are represented as a Directed Acyclic Graph (DAG) $G=(\mathcal{V},\mathcal{E})$, with scenes $S \subset \mathcal{V}$ as leaf nodes. To account for multiple inheritance, we derive semantic similarity from node Information Content, $\mathrm{IC}(u)=1-\frac{\log|\mathcal{D}(u)|}{\log|S|},$
where $\mathcal{D}(u)$ denotes the descendant leaf scenes of node $u$. For a scene pair $(s_i,s_j)$, we identify their Most Informative Common Ancestor ($\mathrm{MICA}$) and define:
\begin{equation}
A_T(s_i, s_j) = \frac{\mathrm{IC}(\mathrm{MICA}(s_i,s_j))}{\max_{u \in \mathcal{V}} \mathrm{IC}(u)}.
\end{equation}

\subsection{Similarity matrix construction and taxonomy comparison}
\label{sec:similarity_matrix_construction}

The scene-level representations define four symmetric similarity matrices:
\begin{align}
    A^{P}_{ij} &= \cos(\mathbf{p}_{s_i}, \mathbf{p}_{s_j}), &
    A^{F}_{ij} &= \cos(\mathbf{f}_{s_i}, \mathbf{f}_{s_j}), \\
    A^{L}_{ij} &= \cos(\mathbf{b}^{\mathrm{raw}}_{s_i}, \mathbf{b}^{\mathrm{raw}}_{s_j}), &
    A^{E}_{ij} &= \cos(\mathbf{b}^{\mathrm{prompt}}_{s_i}, \mathbf{b}^{\mathrm{prompt}}_{s_j}).
\end{align}
We denote these matrices as $A_P$, $A_F$, $A_{L}$, and $A_{E}$ respectively, and compare them with the manual taxonomy view $A_T$. Each similarity matrix is converted to a dissimilarity matrix, $D=1-A_x$ for $x \in \{P,F,L,E,T\}$ for Representational Similarity Analysis (RSA) and hierarchical agglomerative clustering.

\noindent\textbf{Neighborhood overlap.} To capture scene-level differences between acoustic and semantic representations, we compute the Rank-Biased Overlap ($\mathrm{RBO}$) between the top-$k$ nearest neighbors of each scene in acoustic space ($N_A$) and semantic space ($N_T$):
\begin{equation}
\mathrm{RBO}_k(s) = (1-\rho_{\mathrm{RBO}})\sum_{d=1}^{k} \rho_{\mathrm{RBO}}^{\,d-1}\, \frac{|N_A^{1:d}(s) \cap N_T^{1:d}(s)|}{d},
\end{equation}
where $\rho_{\mathrm{RBO}} \in (0,1)$ controls the weighting of higher-ranked neighbors. We report the complementary \emph{neighborhood instability}, $\omega_k(s)=1-\mathrm{RBO}_k(s)$, where larger values indicate stronger acoustic--semantic mismatch.

\subsection{Event-based Naive Bayes classification}
\label{sec:naive_bayes_classifier}
To assess the impact of the acoustic-semantic representations on scene classification, we use an event-based Naive Bayes classifier. For a test clip, the classifier predicts
\begin{equation}
\hat{s} = \arg\max_{s \in S} P(s) \prod_{e \in \mathcal{E}} p(e \mid s)^{x_e},
\end{equation}
where $x_e \in \{0,1\}$ indicates whether event $e$ is present in the clip. We compare two variants that differ only in the event likelihoods: \textit{\textbf{NB Acoustic}} uses the conditional probabilities $p(e \mid s)$ (Acoustic Profile), whereas \textit{\textbf{NB Semantic}} uses the normalized $\mathrm{dPPMI}$-weighted likelihoods $f_s(e)$ (Semantic Profile).

\section{Experimental Setup}
\textbf{Corpus statistics.} The corpus contains \textit{207,315} clip-level scene annotations. Approximately \textit{69\%} correspond to identifiable scenes, including over \textit{130,000} single-label and \textit{12,000} multi-label annotations, while the remaining \textit{31\%} are labeled \textit{unknown}. The Label confidence distribution is approximately as follows: \textit{37\%} of the instances are categorized as \textit{very high} confidence, \textit{28\%} as \textit{high}, \textit{28\%} as \textit{medium}, \textit{6\%} as \textit{low}, and \textit{1\%} as \textit{very low}. For the subsequent analysis, we restrict the corpus to instances with confidence levels ranging from \textit{medium} to \textit{very high}.

\noindent\textbf{Taxonomy structure.} The proposed audiovisual (AV) taxonomy covers \textit{186} standard scenes, and features \textit{12} hierarchy roots, \textit{137} leaf nodes, and a maximum depth of four. The audio-centric (AC) taxonomy includes \textit{75} standard scenes, and contains \textit{9} hierarchy roots, \textit{52} leaf nodes, and a maximum depth of four. For the subset selected based on label confidence, the AV taxonomy has \textit{135} scene labels; the AC taxonomy, \textit{68}. Standardization of scene labels is made with \texttt{py-salt} \cite{Stamatiadis2024}. 

\noindent\textbf{Pipeline quality validation.} To assess label quality, we manually annotated a subset of the AudioSet Strong evaluation set comprising \textit{1,284} clips. Two independent annotators labeled each clip under both taxonomies, reaching substantial agreement (Cohen's $\kappa=0.634$ for the audiovisual taxonomy, $\kappa=0.611$ for the audio-centric taxonomy).

\noindent\textbf{Classifier probing.} Naive Bayes with Laplace smoothing $\alpha=1.0$; performance reported as accuracy and Macro-F1 score to account for long-tail class imbalance on the manually annotated test set.

\noindent\textbf{Neighborhood evaluation.} We report scene neighborhood instability $\omega_k$ over $k = 1, \ldots, 25$, with RBO decay $\rho_{\mathrm{RBO}}=0.90$. 

\section{Results and Discussion}
\label{sec:results}

\subsection{Alignment of scene representations with the human taxonomy}

\begin{table}[t]
\centering
\caption{Mantel test Spearman correlations ($r_M$) between representational similarity matrices across views and taxonomies. Audiovisual (AV) and audio-centric (AC) results are reported. All correlations are significant ($p=0.0001$).}
\label{tab:mantel_matrix}

\small
\setlength{\tabcolsep}{4pt}
\begin{tabular}{llcccc}
\toprule
\textbf{View} & \textbf{Tax.} & Acoustic & Semantic & Lexical ($A_L$) & Lexical ($A_E$) \\
\midrule

\multirow{2}{*}{Manual}
& AV & 0.123 & 0.312 & 0.326 & 0.253 \\
& \cellcolor{gray!10}AC 
& \cellcolor{gray!10}0.183 
& \cellcolor{gray!10}0.336 
& \cellcolor{gray!10}0.277 
& \cellcolor{gray!10}0.192 \\

\midrule

\multirow{2}{*}{Acoustic}
& AV &  &  & 0.167 & 0.231 \\
& \cellcolor{gray!10}AC 
& \cellcolor{gray!10} 
& \cellcolor{gray!10} 
& \cellcolor{gray!10}0.225 
& \cellcolor{gray!10}0.258 \\

\midrule

\multirow{2}{*}{Semantic}
& AV &  &  & 0.313 & 0.285 \\
& \cellcolor{gray!10}AC 
& \cellcolor{gray!10} 
& \cellcolor{gray!10} 
& \cellcolor{gray!10}0.390 
& \cellcolor{gray!10}0.345 \\

\bottomrule
\end{tabular}
\end{table}

\label{sec:rsa_results}
We assess how well each view recovers human scene organization with RSA, computing Mantel-tested Spearman correlations between the dissimilarity matrices of each view and of the manual taxonomy $A_T$ over the AV and AC scene pairs (all p = 0.0001).

\Cref{tab:mantel_matrix} reveals a clear abstraction gradient. The acoustic profile aligns weakly with the human taxonomy ($0.123$ AV; $0.183$ AC), whereas the semantic profile improves it substantially under both taxonomies ($0.312$ AV; $0.336$ AC), indicating that human scene organization aligns more with discriminative event-scene associations.

The lexical views give a complementary reading. The semantic profile is closer than the acoustic profile to both lexical views, but the gap shrinks for $A_E$ ($0.285$ vs. $0.231$ AV; $0.345$ vs. $0.258$ AC, against $0.313$ vs. $0.167$ AV and $0.390$ vs. $0.225$ AC for $A_L$). The acoustic profile thus behaves more like a language representation of \textit{what a place sounds like}, whereas the semantic profile is closer to the abstract, label-level notion of the scene. With respect to human taxonomy - lexical alignment, $A_T$, $A_L$ is highest in AV ($0.326$ vs. $0.312$ for the semantic profile), while in AC the semantic profile is highest overall ($0.336$ vs. $0.277$). Prompting lowers taxonomy alignment ($A_E$: $0.253$ AV; $0.192$ AC), suggesting that sound-oriented descriptions move away from the abstract taxonomy, though this may depend on the template. Fig. 3 illustrates these neighborhood differences.


\subsection{Scene-level neighborhood instability}
\label{sec:overlap_results}

\begin{figure}[t]
  \centering
tt  \centerline{\includegraphics[width=\columnwidth]{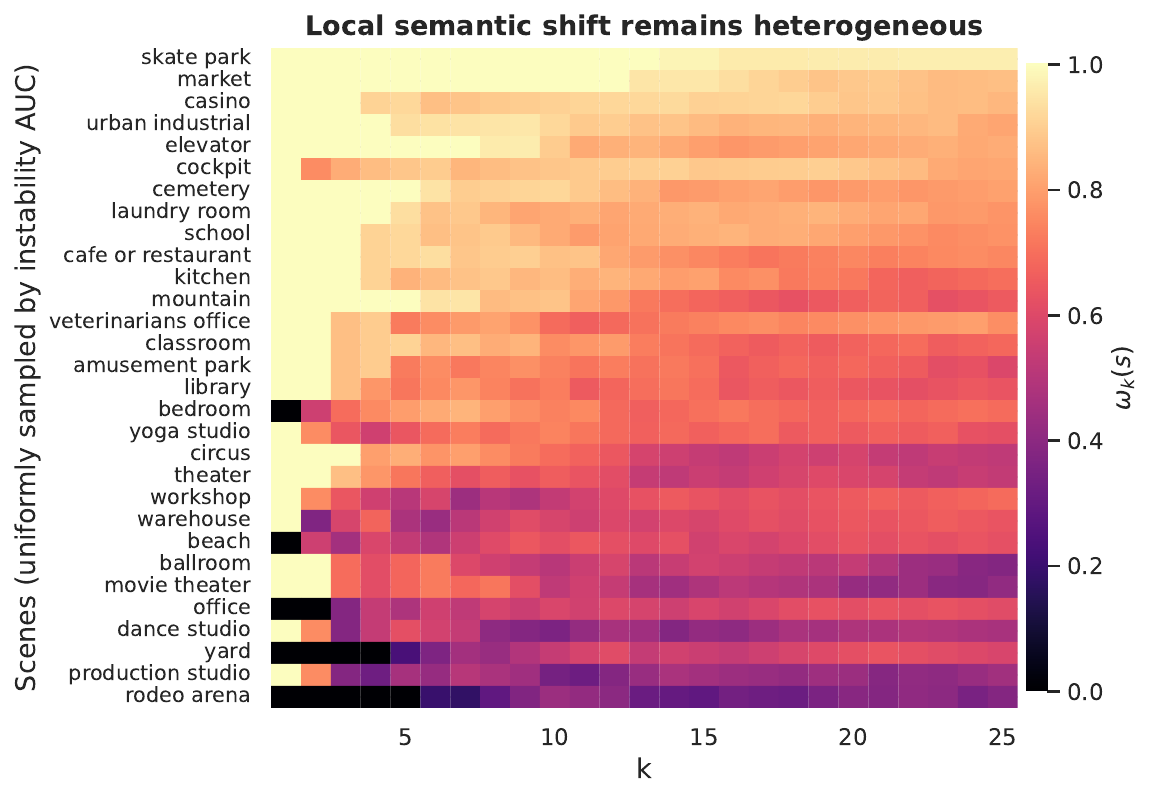}}
    \caption{Heatmap of scene-level semantic shift, $\omega_k(s)=1-\mathrm{RBO}_k(s)$, across neighborhood scales ($k=1,\ldots,25$) for the audiovisual taxonomy. Scenes are ordered by instability; higher = greater acoustic--semantic mismatch.}

  \label{fig:heatmap}
\end{figure}

The RSA correlations in the previous section summarize global alignment, and can hide substantial scene-level variability. We now analyze the neighborhood instability $\omega_k(s)$, introduced in \cref{sec:similarity_matrix_construction}, which measures the divergence between the top-$k$ acoustic and semantic neighbors of each scene. Ranking scenes by the AUC of $\omega_k$ over increasing neighborhoods (\cref{fig:heatmap}) reveals two regimes. Unstable scenes such as \textit{skate park}, \textit{market}, and \textit{casino} exhibit persistent acoustic--semantic neighborhood mismatch, as their dominant sounds are shared across diverse environments. In contrast, distinctive scenes such as \textit{rodeo arena}, \textit{dance studio}, and \textit{office} remain stable because their acoustic neighborhoods already reflect their semantic identity.

The analysis further reveals two complementary behaviors. \textit{\textbf{Semantic anchors}} such as \textit{production studio} and \textit{yard} show consistently low $\omega_k$, indicating that acoustic and semantic neighborhoods are already aligned. Conversely, \textit{\textbf{acoustic chameleons}} such as \textit{skate park} and \textit{market} exhibit high instability: their raw acoustic profiles are dominated by shared background sounds, while the semantic representation shifts neighborhoods toward more conceptually related scenes. These mismatches explain why global alignment remains moderate despite strong semantic structure in the proposed representation.

\subsection{Downstream classification analysis}
\label{sec:nb_results}

To test whether the observed geometric differences relate to classification performance, we evaluate the two Naive Bayes classifiers from \cref{sec:naive_bayes_classifier} under both taxonomies (\cref{tab:nb_scene_classification}). Accuracy remains comparable and slightly favors the \textit{NB Acoustic} classifier, whereas the \textit{NB Semantic} classifier achieves higher Macro-F1 (0.1282$\rightarrow$0.1315 AV; 0.1847$\rightarrow$0.1860 AC). This suggests that discriminative event weighting benefits lower-support, confusable scenes, improving class-balanced performance with little loss in accuracy.

\begin{figure}[t]
  \centering
  \centerline{\includegraphics[width=\columnwidth]{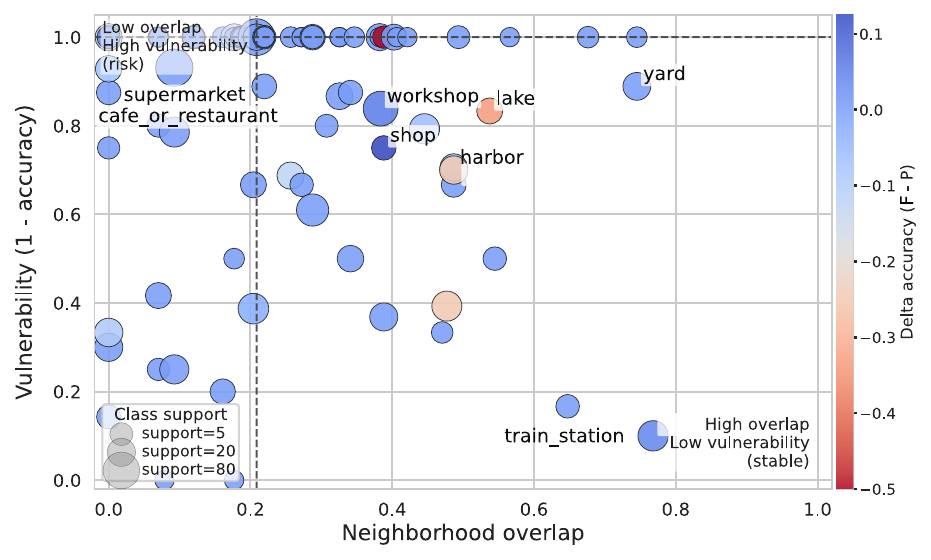}}
  \caption{Intuition map relating neighborhood overlap and Naive Bayes vulnerability. It indicates stable and vulnerable regimes.}

  \label{fig:overlap_nb}
\end{figure}

\Cref{fig:overlap_nb} links neighborhood overlap to the accuracy difference between \textit{NB Semantic} and \textit{NB Acoustic} classifiers for each scene. Two trends stand out. Firstly, scene-level trends echo the neighborhood instability patterns in \cref{sec:overlap_results}. Acoustic chameleons such as \textit{workshop} and \textit{shop} improve under the \textit{NB Semantic} classifier, suggesting that semantic reweighting can unmask scene-specific cues by down-weighting broadly shared events. Conversely, scenes such as \textit{harbor} and \textit{lake} may degrade when useful but non-exclusive acoustic evidence is suppressed. Secondly, scenes with lower acoustic--semantic overlap also tend to be more sensitive to semantic reweighting, suggesting that $\omega_k$ provides an indicator of potential classification difficulty. This relationship is not deterministic. \textit{train station} remains stable, with high overlap and little performance change, whereas \textit{supermarket} and \textit{cafe or restaurant} exhibit both low overlap and high vulnerability. Overall, neighborhood overlap provides a useful indicator of when the relative contributions of acoustic and semantic evidence are likely to matter. Combined with class-specific event profiles, it suggests a basis for adaptive classifiers that rely primarily on acoustic evidence while selectively incorporating discriminative semantic weighting for scenes with greater contextual ambiguity.



\begin{table}[t]
\centering
\caption{Scene classification performance of the Naive Bayes classifiers based on acoustic and semantic views under the AV and AC evaluation taxonomies. Classifiers receive event labels as input and predict the corresponding scene.}
\label{tab:nb_scene_classification}

\sisetup{
    reset-text-series = false,
    text-series-to-math = true,
    mode=text,
    round-mode=places,
    round-precision=4,
    table-format=1.4,
    table-number-alignment=center
}

\begin{tabular}{llSS}
\toprule
\textbf{Evaluation taxonomy} &
\textbf{Classifier} &
{\textbf{Accuracy} $\uparrow$} &
{\textbf{Macro F1} $\uparrow$} \\
\midrule
\multirow{2}{*}{Audiovisual (AV)}
    & NB Acoustic & \bfseries 0.1881 & 0.1282 \\
    & NB Semantic & 0.1787 & \bfseries 0.1315 \\
\midrule
\multirow{2}{*}{Audio-centric (AC)}
    & NB Acoustic &  \bfseries 0.2816 & 0.1847 \\
    & NB Semantic & 0.2576 & \bfseries 0.1860 \\
\bottomrule
\end{tabular}
\end{table}

\section{Conclusion}

We extended AudioSet Strong with clip-level scene annotations under two complementary taxonomies. Using this augmented resource, we showed through a multi-view acoustic-semantic analysis that raw event frequency only weakly reflects human scene organization, whereas weighted associations recover substantially more of this structure. Neighborhood instability distinguishes semantic anchors from acoustic chameleons and identifies scenes sensitive to event reweighting. Overall, event semantics provides a stronger link to scene identity than acoustic profiling, while neighborhood instability offers an unsupervised signal for anticipating where joint scene–event models are most likely to struggle.

\section{Acknowledgment}
This work was supported by Hi!Paris, State funding managed by the French National Research Agency (ANR) under the France 2030 program, reference ANR-23-IACL-0005, and by the European Union (ERC, HI-Audio, 101052978). Views and opinions expressed are however those of the author(s) only and do not necessarily reflect those of the European Union or the European Research Council. Neither the European Union nor the granting authority can be held responsible for them.


\bibliographystyle{IEEEtran}
\bibliography{refs}







\end{document}